# Influência de fatores organizacionais e sociais na etapa de levantamento de requisitos

GLAUBER DA ROCHA BALTHAZAR
Instituto de Pesquisas Tecnológicas – São Paulo – Brasil
glauber_rochab@yahoo.com.br

MARCIA ITO
Laboratório de Pesquisa em Ciências de Serviço – Programa de Mestrado –
Centro Estadual de Educação Tecnológica "Paula Souza" (CEETEPS)
– São Paulo – SP – Brasil
ito@mind-tech.com.br

**Resumo –** A etapa mais crítica e frágil do projeto de construção de um software é a de levantamento de requisitos e, por conta disso, a Engenharia de Requisitos vem evoluindo suas técnicas no intuito de minimizar ao máximo os problemas que o Analista de Requisitos enfrenta. Porém, poucos trabalhos levam em consideração as relações humanísticas e os comportamentos dos envolvidos nesta etapa. Este artigo apresenta um levantamento de alguns trabalhos realizados nesta etapa que levam em consideração fatores não técnicos como emoções, ambiente organizacional e social dos envolvidos.

<u>Palavras-chave</u>: Engenharia de Requisitos, Emoções em Requisitos, Resistência à mudança.

**Introdução**

O período definido pela etapa de levantamento das necessidades dos usuários para a construção de um software é caracterizado por uma forte interação social entre os Analistas de Requisitos e os usuários envolvidos. Além disso, esta fase também é marcada por aspectos que tem características não apenas técnicas como fatores organizacionais e sociais [1]. Desta forma, alguns autores como Bergman [2] e Ramos [3] citam em seus trabalhos que apenas os avanços nas tecnologias e técnicas de modelagem por si só não são suficientes para proteger projetos grandes e complexos de falhas.

Assim, estes trabalhos procuram apresentar que muitos dos fatores organizacionais e sociais podem influenciar na elicitação de requisitos. Estes fatores são definidos como "sentimentos negativos que não são adaptados à realidade do trabalho" [3] e, além disso, associado a estes fatores têm-se um ambiente organizacional muito complexo. Desta forma, o Analista de Requisitos além de compreender as metodologias, processos e técnicas existentes na Engenharia de Software deve também tentar compreender como as mudanças tecnológicas, organizacionais, humanísticas e institucionais estão intrinsecamente interligadas com as falhas que ocorrem nos projetos de software [2]. Esta compreensão tem como objetivo conseguir interpretar melhor os usuários com os quais ele irá interagir e assim conseguir criar tanto um ambiente social mais harmônico quanto obter a confiança dos usuários para que nenhuma informação seja mal informada ou esquecida.

Este artigo tem como objetivo apresentar os estudos existentes que procuram conceituar os aspectos sociais e organizacionais que podem influenciar na etapa de levantamento de requisitos. Assim serão apresentados alguns trabalhos que definem como as emoções das pessoas são expostas, como elas reagem as transformações e mudanças, quais são seus valores e crenças perante as mudanças, se elas são influenciadas por gerentes e superiores existentes no seu trabalho e como a comunicação entre elas irá servir como meio para a propagação destes elementos.

Assim, este trabalho está organizado da seguinte forma: a seção seguinte (Complexidade na Engenharia de Requisitos) apresenta uma breve revisão dos trabalhos que já foram feitos procurando compreender os aspectos organizacionais e sociais que influenciam no processo de requisitos de modo a servir de base para a fundamentação do levantamento apresentado neste artigo. Em seguida são apresentados diversos trabalhos que podem ser feitos além dos existentes como complementos para estas pesquisas.

## Complexidade na Engenharia de Requisitos

A expressão "complexidade no levantamento de requisitos" foi apresentada por Ramos [3] como sendo um "conjunto de elementos humanísticos que ocorrem na fase de levantamento de requisitos". Porém, para este trabalho estende-se esta definição de modo a complementá-la dizendo que estes elementos não estão diretamente relacionados aos elementos trabalhados nas técnicas, processos e metodologias atualmente existentes na Engenharia de Software. Como consequência, abre-se um leque de oportunidades de pesquisas que possam levar em consideração outros elementos como emoção [3], resistência a transformação [3], valores e crenças dos usuários e analistas [3], influência de gestores [4], problemas na gestão de projetos [3] [5], ruídos na comunicação [6], negociação de conflitos [6], etnografia [7] e processo grupal [8] [9]. Estes elementos são descritos a seguir.

## Emoção

Os aspectos emocionais são comportamentos que os usuários expressam diante do software e que, geralmente, não são levados em consideração na Especificação dos Requisitos. Estes comportamentos são caracterizados como sensações, comportamentos e atitudes que os usuários expressam quando expostos ao sistema já informatizado (ou informatização de um sistema) que podem ser, por exemplo, tanto um medo diante da adoção de um novo sistema e a mudança da sua forma de trabalho quanto uma vergonha de um usuário em utilizar perante seus colegas um software com uma interface mais feminina (com cores rosa, por exemplo).

O trabalho, realizado por Barros [10] e intitulado como Requisitos Emocionais em Jogos apresenta a concepção de incorporar na Especificação de Requisitos os fatores emocionais. Neste trabalho ele apresenta a ideia de que, no desenvolvimento de softwares para jogos deve-se levar em consideração não apenas os requisitos tradicionais (explícitos como requisitos funcionais e não funcionais), mas sim que é "preciso analisar características emocionais dos usuários; procurar entender como o público-alvo deve reagir ao programa; e que sensações este software será capaz de proporcionar aos seus usuários". Desta forma, ele sugere que além dos requisitos tradicionais também deve ser levado em consideração as características emocionais dos usuários que irão interagir com o software.

Apesar de Barros [10] citar que as emoções devem ser consideradas, ele afirma que isso não ocorre com frequência nas especificações de requisitos. Ele diz que isso pode ser recorrente da subjetividade que as emoções proporcionam e, desta forma, não fica claro como descrevê-las. Isso ocorre até em áreas em que o tema já é abordado há algum tempo, como a área de Interação Humano-Computador (IHC) tendo em vista que ela tem o objetivo de permanecer direcionada "na observação e construção de medições específicas de comportamentos apresentados por usuários de determinados produtos" [10] [11].

**Resistência à transformação**

A resistência das pessoas a mudanças é uma forma de reação de controle que se observa dos seus comportamentos com as ações do meio a que elas estão sujeitas. Esta resistência trata-se de uma classe de comportamentos sociais que seguem as mesmas leis que regem qualquer outro comportamento humano. Porém, ocorre que as pessoas já têm um conjunto de hipóteses pessoais (constituídas por mitos e valores) que caracterizam um arcabouço de ideias, pensamentos, crenças e pressupostos e estas hipóteses entram em conflito com as hipóteses propostas nas transformações. Como consequência, realiza-se o início da resistência às mudanças, pois, estas resistências deixam de ter um enunciado teórico e geral ("mudanças precisam ocorrer na minha vida...", por exemplo) e passam a ser explícitas e específicas ("você precisa aprender a nova forma de trabalho!", por exemplo) [12].

Ramos [13] publicou um trabalho denominado *Requirements Engineering for Organizational Transformation* que faz uma referência direta as resistências à transformação que os usuários estão sujeitos dentro de um contexto organizacional. Neste trabalho ela considera que a resistência a mudanças pode ocorrer quando se é introduzido um Sistema Baseado em Computação (SBC) para automatizar um processo de trabalho. Assim, devido ao fato de que um SBC transforma a organização, alterando os padrões e formas de trabalho dos usuários neste meio, a resistência nasce por conta de que estas mudanças interagem diretamente com os valores e crenças dos usuários e, como resultado, desencadeiam em respostas emocionais que são por vezes dirigidos contra o SBC e os seus responsáveis [13].

Ramos [13] defende a ideia de que a resistência à transformação é um fato que já deverá ser esperado quando se introduz um SBC e que, além disso, promove também uma transformação na organização. Esta transformação é caracterizada tanto como uma nova forma de trabalho quanto uma nova lógica no processo de trabalho. Logo, esta resistência deve ser vista como um problema prévio a ser resolvido (ou minimizado) o mais rápido possível. Na prática as pessoas devem ser motivadas para reagir a estas mudanças de forma colaborativas e flexíveis. Para tanto, as transformações das organizações devem ser transmitidas para os funcionários através da adoção de técnicas (palestras ou conversas individuais) que demonstram como estas mudanças são positivas, pois irão melhorar a produção, a qualidade do trabalho, a eficiência e eficácia organizacional reduzindo assim custos, melhorando o desempenho, tanto individual quanto em grupo, e ainda capacitar as pessoas para explorar o máximo do seu potencial.

**Valores e Crenças**

Valores e crenças são conceitos sociais que definem as percepções de certo e errado que são utilizados para julgar os conceitos de bom ou ruim nos resultados das ações que as pessoas tomam. Além disso, estes conceitos também são as percepções que as pessoas detêm sobre a realidade em que elas vivem e isso ocorre porque estas pessoas querem definir sua posição e determinar as possíveis relações que elas podem interagir com as partes que compõem esta realidade [3] [14]. Estas percepções têm um caráter dinâmico e de contínua transformação se forem construídos a partir das histórias de vida e das múltiplas aprendizagens e experiências vivenciadas pelo sujeito mediante suas interações com os diferentes aspectos do contexto histórico sociocultural [15]. É importante mencionar que se por um lado a influência sociocultural é heterogênea (uma grande diversidade cultural) e muitas vezes contraditória (possibilitando constantes processos de mudança dentro do sistema de orientações para crença), por outro lado, a dimensão afetiva no mundo psicológico do indivíduo leva a uma maior constância ou estabilidade nesse sistema, o que, assim, caracteriza o conceito de valor para o sujeito. Por fim, "dependendo do significado e da intensidade das experiências nas quais as crenças são construídas, estas serão mais ou menos resistentes a mudanças, gerando uma maior ou menor transformação no sistema de crenças e valores do indivíduo" [15].

O trabalho realizado por Ramos [3], denominado *The Role of Emotion, Values, and Beliefs in the Construction of Innovative Work Realities* tem como objetivo observar como os sentimentos de valores e crenças influenciam em uma Transformação Organizacional ocorrida pela implantação de um software. Esta transformação nada mais é que o processo de alteração da forma de trabalho de uma organização quando esta decide automatizar os seus processos e percebe que a automatização irá alterar a forma de trabalho existente e que isso será benéfico para a organização.

Porém, em uma Transformação Organizacional conceitos e práticas muitas vezes levam a conflitos com as crenças e valores dos indivíduos envolvidos. Ou seja, uma mudança que for tão radical na forma em que os sujeitos envolvidos compreendem a realidade e os seus papéis e atividades, poderá desencadear diversas emoções sociais que precisam ser cuidadosamente tratadas, de modo paciente, ao longo do tempo. Para exemplificar, Isabel Ramos [13] apresenta no seu trabalho uma breve estória na qual ela mostra um caso simples em que uma determinada pessoa não gosta da cor azul. Ela cita que ninguém pode forçar esta pessoa a gostar, porém ela pode ser instruída a demonstrar algum sentimento favorável para gostar desta cor, porém isto será maléfico, pois somente irá aumentar o seu desagrado. Esta pessoa então poderia ser motivada a gostar da cor azul pelas vantagens dela. Isto não significará que ela irá mudar as suas preferências de cor. Assim se faz necessário um esforço para que uma pessoa aceite uma mudança devido ao fato de ela já ter os valores formados quando deparado com algo que ela julga pejorativo e que isso ainda é agravado pelo fato de que

o ser humano é resistente a mudanças [3].

**Problemas de gestão de projetos**

Gestão de projetos é uma disciplina responsável por estabelecer os objetivos e parâmetros a serem seguidos no desenvolvimento de projetos, objetivando definir os escopos segundo especialidades e etapas. Além disso, cabe a esta disciplina planejar os custos de desenvolvimento dos projetos, planejar as etapas e prazos de desenvolvimento das diversas etapas tanto no todo do projeto quando nas suas especialidades. A gestão também é responsável por garantir a qualidade das soluções técnicas adotadas validando as etapas e fomentar a comunicação entre os participantes, coordenando as interfaces e garantindo a compatibilidade entre as soluções [16].

Esta sub seção apresenta dois trabalhos (Ramos [13] intitulado *Requirements Engineering for Organizational Transformation* e Candida [5] intitulado "Introdução e uma visão do processo de software") que deduzem a hipótese de que problemas na gestão de projetos influenciam nos requisitos. Alguns destes problemas são mostrados por Candida [5] quando ela afirma que os gerentes são responsáveis por cobrar e exigir estimativas de prazos dos Engenheiros de Software e esta cobrança as vezes é feita de forma tão intensa que, devido à pressão, os engenheiros tendem a fornecer prazos que não podem ser cumpridos. Assim, prazos irreais são definidos e um novo termo surge, denominado "prazo político" que nada mais é que aquele prazo que é informado apenas para responder a pressão da solicitação e não condiz com a realidade. Como consequência, para os desenvolvedores significa o surgimento de stress e má qualidade de vida no trabalho e para os gerentes a perda de credibilidade e prejuízos. Além disso, como produtos finais são gerados artefatos de má qualidade e mais caros que deveriam (além, de serem entregues fora do prazo).

Ramos [13] considera no seu trabalho que os problemas de gestão podem surgir de atitudes emotivas que poderiam ser resolvidas com uma simples mudança de funcionalidade ou forma de abordagem do usuário. Nesta visão, a maioria dos problemas apresentados por Candida [5] podem se tornar problemas de requisitos porque irão refletir no desenvolvimento ou na implantação de um sistema. De uma forma geral, qualquer problema que possa impedir o êxito de implantação de um sistema ou a devida compreensão das funcionalidades devem ser identificados o mais cedo possível para evitar que ele possa permear no transcorrer do processo de execução do projeto [13].

**Influência dos gestores**

Al-Rawas [4], em seu artigo denominado "*Communication Problems in Requirements Engineering: a Field Study*" realizou uma pesquisa sobre como é feita a seleção dos usuários que irão participar no processo de levantamento de requisitos. Neste contexto ele aplicou diversas pesquisas e realizou algumas observações de modo a determinar como se dão estas escolhas além de descrever como é o perfil dos usuários envolvidos e até quanto a escolha é influenciada por aspectos que não são definidos exclusivamente pelo Analista de Requisitos.

Assim, Al-Rawas [4] conseguiu perceber que alguns fatores eram levados em consideração pelos analistas de requisitos na escolha dos usuários elegíveis ao processo. Os fatores utilizados são:

- recomendações dos gestores: os gestores indicavam funcionários de sua empresa para participarem deste processo,
- conhecimento dos usuários do negócio: aqueles usuários que estavam envolvidos diretamente com a execução prática dos processos do negócio, que dominavam o "como é feito",
- conhecimento dos usuários de informática: usuários que independente da sua relação com o negócio, tanto prática quanto gerencial, tinham um bom conhecimento de informática e até de construção de software e
- nível organizacional do usuário: a escolha era feita no nível funcional do usuário, por exemplo, líderes, gestores ou supervisores.

Baseado nestes fatores, Al-Rawas [4] realizou uma pesquisa e questionou um conjunto de analistas para que eles escolhessem o fator que mais influenciava a sua escolha para a seleção de usuários que iriam participar das reuniões de levantamento de requisitos. O resultado foi que grande parte dos analistas diziam que "o conhecimento dos usuários do negócio" era o fator mais relevante perante os demais.

Porém, apesar dos analistas terem respondido nesta pesquisa que o conhecimento do negócio era o mais importante, foi percebido que a escolha era também fortemente influenciada pelas recomendações dos gestores, pois estes acabavam sugerindo outras pessoas para participarem desta fase. Assim, os gestores acabavam influenciando na escolha com a argumentação de tentar encontrar um equilíbrio dos envolvidos pelo fato de os analistas não escolherem gerentes e gestores que estavam voltados apenas para a gerencia da empresa. O objetivo desta influencia era permitir que os profissionais de software conseguissem se comunicar tanto com as pessoas envolvidas no negócio quanto com aquelas que estavam mais relacionadas a gerencia da empresa e, como consequência, terem uma visão ampla do funcionamento de todo o negócio.

Contudo, na prática, ficou evidenciado que estes usuários (que os gestores envolviam no processo de levantamento) não tinham conhecimento das informações que realmente eram importantes para o analista de requisitos. Isto fica evidente através da pesquisa aplicada nos usuários, sobre qual o impacto efetivo da interferência deles na construção do software. Nesta pesquisa os usuários envolvidos foram convidados a responderem qual era realmente a sua interferência no processo de trabalho que seria automatizado.

**Comunicação**

A fase de levantamento de requisitos é caracterizada pela intensidade e importância das atividades de comunicação. Durante esta fase, as diversas partes interessadas devem ser capazes de comunicar as suas necessidades com os analistas e estes devem ser capazes de comunicar as especificações que eles geram para as partes interessadas objetivando realizar validações [4]. Desta forma, esta comunicação compreende um estágio contínuo e altamente crítico para o desenvolvimento do sistema e, além disso, está sujeito a um elevado grau de erros, pois são fortemente influenciados por problemas de comunicação [6].

Para Passadori [17] estes problemas de comunicação são categorizados

em três grupos:
- Psicológicos: compreendem as sensações e sentimentos dos envolvidos como, por exemplo, medo, excesso de preocupação e baixa auto-estima;
- Físicos: compreendem problemas relacionados a parte física dos envolvidos como, por exemplo, voz fraca, dicção ruim, velocidade excessiva ou demasiadamente lenta da voz, nasalação, gestos excessivos e postura inadequada; e
- Técnicos: aqueles de origem da atuação dos profissionais em suas atividades como, por exemplo, desorganização de ideias, vícios de linguagens, dificuldade de vocabulário técnico e dificuldade com as normas da língua para escrita de textos.

Complementando Al-Rawas [4] afirma que os problemas de comunicação ocorrem por conta de que os Analistas de Requisitos não têm todos os conhecimentos necessários para o projeto e, por conta disso, eles devem adquirir informações adicionais (com os interessados pelo sistema) antes de realizar o trabalho produtivo. Assim, a presença destes problemas influenciam diretamente tanto na qualidade da aquisição quanto na partilha das informações entre os vários membros constituintes do projeto.

Baseado nesta falta de conhecimento e, como consequência, da necessidade de uma boa comunicação, Faulk [18] organiza e detalha o problema de comunicação em uma categoria denominada "problema essencial". Nesta categoria ele cita que os elementos que a caracterizam são inerentes à elicitação de requisitos como: dificuldades do usuário em saber efetivamente o que ele quer e dificuldade de comunicação entre usuário e desenvolvedor por conta da natureza mutante dos requisitos. Além disso, os problemas essenciais ocorrem por conta de dois fatores primordiais:
a. Compreensão: as pessoas não sabem exatamente o que querem; e
b. Comunicação: requisitos de software são difíceis para se comunicar efetivamente (principalmente os requisitos não funcionais ).

Enfim, para Faulk [18] "a dificuldade inerente da comunicação é composta pela diversidade de pessoas e audiências para a especificação de requisitos".

**Negociação de conflitos**

Conflitos são partes constituintes e inevitáveis da vida quotidiana do ser humano e, além disso, estão presentes tanto nas esferas da vida humana (psicológica, política, econômica, religiosa, social e cultural) quanto entre todos os tipos de relações existentes (interpessoais, conjugais, trabalhistas, étnicas e internacionais). Apesar do conceito de conflito ser definido de diferentes formas por diferentes autores (por conta das diferentes épocas da sociedade) a definição descrita pelo alemão Georg Simmel, de acordo com Bartholo [19], é vista como a que melhor se encaixa na atual sociedade. Assim, os conflitos são formas de interação social e, portanto, constituintes das relações sociais. Assim, os conflitos não estão apenas presentes na sociedade mas são indispensáveis e desempenham o importante papel de solucionar dualismos divergentes, ou seja, é "um modo de conseguir algum tipo de unidade, ainda que através da aniquilação de uma das partes conflitantes" [19]. Conclui-se então que os conflitos são meios pelos quais os atores sociais extinguem suas divergências, interesses antagônicos ou pontos de vista, possibilitando que a sociedade alcance certa unidade [19].

Utilizando o conceito de que conflitos são formas de interações sociais e baseado na afirmação de Bartholo [19] que diz que na Engenharia de Requisitos a atividade mais intensa é a de interação entre pessoas (analistas e usuários), [6], em seu trabalho intitulado *Communication Issues in Requirements Elicitation: A Content Analysis of Stakeholder Experiences*, afirma que os conflitos no contexto do levantamento de requisitos ficam evidentes quando as pessoas envolvidas adotam papéis nos relacionamentos entre si. Esta adoção de papéis tem como consequência o nascimento da figura de moderador e esta figura fornece aos envolvidos um método pelo qual o conhecimento será explorado e, como consequência, será determinada uma forma de negociação entre os possíveis conflitos que surgirão.

Nesta interação, as relações podem revelar-se problemáticas dada a divisão do conhecimento especialmente entre os negócios e as bases de Tecnologia da Informação especialmente no alinhamento das informações dos interessados. Isso ocorre devido ao fato de que as considerações técnicas e sociais tem o mesmo peso para ambos os envolvidos. Assim o papel de mediador será o responsável por determinar a forma como estas informações serão trocadas, ou seja, o mediador irá determinar um entendimento do problema e as possíveis soluções [6].

Para tanto, o mediador deverá se utilizar de um modelo de interação para determinar como mediar o entendimento do problema com as possíveis soluções conflitantes. Segundo Coughlan [6] dois modelos são descritos para isso, sendo eles: modelo de integração do conhecimento e modelo de aprendizagem mútua.

O modelo de integração do conhecimento é focado na comunicação entre usuários e desenvolvedores de sistema. Neste modelo são destacados alguns fatores importantes como: resultados do processo de desenvolvimento do sistema; pré-requisitos dos solicitantes e ferramentas e técnicas para a descrição do sistema. Porém, o modelo foca todos os fatores na distinção de três domínios do discurso. O segundo modelo, modelo de aprendizagem mútua, Bartholo [19] apresenta uma forma de manter uma relação mais estreita entre analistas e usuários na qual é sugerida uma grande relação de colaboração entre estes dois atores. Neste contexto Bartholo [19] expressa que os usuários não tem uma boa habilidade para falar de forma eficaz sobre o seu trabalho, porém, eles conseguem explicar muito bem a parte prática, ou seja, apenas como o trabalho é feito e não todos os motivos das regras envolvidas. Assim, o modelo sugere que analistas e usuários tenham conversas sobre a forma de execução do trabalho que irá orientar o analista a compreender como o trabalho é feito.

Por fim, Coughlan [6] conclui que o elo comum entre estes dois modelos está na premissa de que as interações entre as partes interessadas requer a cooperação a fim de desenvolver uma compreensão partilhada da situação em questão.

**Conclusões e trabalhos futuros**

Todos os trabalhos apresentados neste artigo procuram demonstrar que a etapa de levantamento de requisitos apresenta uma complexidade que vai muito além dos métodos, processos e técnicas existentes na Engenharia de Requisitos. Desta forma, deve-se compreender todo o ambiente em que o software será inserido para ter uma melhor visão do impacto que ele irá causar e até que ponto isso irá influenciar nos requisitos. Uma pesquisa a ser sugerida como trabalho futuro envolve descobrir se as interações sociais também influenciam os requisitos. Assim é sugerido que os analistas e usuários venham a ter seus comportamentos estudados e assim será possível determinar se os sentimentos expressos por eles influenciam ou não na escolha dos requisitos para o software.

**Referências**